\documentclass[review]{elsarticle}

\usepackage{epsfig}
\usepackage{amssymb}

\usepackage{amsmath}
\usepackage{amsfonts}
\usepackage{multirow}
\usepackage{graphicx,psfrag,epsf}
\usepackage{enumerate}
\usepackage{natbib}
\usepackage{url} 

\usepackage{lineno,hyperref}
\modulolinenumbers[5]

\journal{Mathematics and Computers in Simulation}

\begin{document}
\begin{frontmatter}

\title{Homogeneity problem for basis expansion of functional data with applications to resistive memories}


\author[mymainaddress]{Ana M. Aguilera\corref{mycorrespondingauthor}}
\address[mymainaddress]{Department of Statistics and O.R. and IEMath-GR, University of Granada, Spain}
\cortext[mycorrespondingauthor]{Corresponding author}
\ead{aaguiler@ugr.es}

\author[mymainaddress]{Christian Acal}

\author[mysecondaryaddress]{M. Carmen Aguilera-Morillo}
\address[mysecondaryaddress]{Department of Statistics. University Carlos III of Madrid, Spain}

\author[mytertiaryaddress]{Francisco Jim\'enez-Molinos}
\address[mytertiaryaddress]{Department of Electronics and Computer Technology. University of Granada, Spain}

\author[mytertiaryaddress]{Juan B. Rold\'an}

\begin{abstract}
The homogeneity  problem for testing if  more than two different samples come from the same population is considered for the case of functional data. The methodological results are motivated by the study of homogeneity of electronic devices fabricated by different materials and active layer thicknesses. In the case of normality distribution of the stochastic processes associated with each sample, this problem is known as Functional ANOVA problem and is reduced to test the equality of the mean group functions (FANOVA). The problem is that the current/voltage curves associated with Resistive Random Access Memories (RRAM) are not generated by a Gaussian process so that a different approach is necessary for testing homogeneity. To solve this problem two different  parametric and nonparametric approaches based on basis expansion of the sample curves are proposed. The first consists of testing multivariate homogeneity tests on a vector of  basis coefficients of the sample curves. The second is based on dimension reduction by using functional principal component analysis of the sample curves (FPCA) and testing multivariate homogeneity on  a vector of principal components scores. Different approximation numerical techniques are employed to adapt the experimental data for the statistical study. An extensive simulation study is developed for analyzing  the performance of both approaches in the parametric and non-parametric cases. Finally, the proposed methodologies are applied on three samples of experimental reset curves measured in three different RRAM technologies.
\end{abstract}


\end{frontmatter}


\section{Introduction}

 The methodological results in this paper linked to homogeneity tests for functional data are motivated by the study of variability in Resistive Random Access Memories. In this work, the devices under study were fabricated making use of different materials for the metal electrodes and dielectrics of different thicknesses. RRAMs are currently considered a serious contender for non-volatile memory applications. These devices operate under the principles of resistive switching (RS), i.e., their internal resistance is switched between different values by changing the nature and features of charge conduction within a dielectric layer. Many different developments are being considered in the research of these devices such as fabrication and characterization; also, the simulation and modeling facets are under study for this emerging technology \cite{G.Cordero16,Villena17}.

The use of devices based on RS for cryptographic applications is based on the inherent stochasticity of their operation. The device resistive state changes in many cases because of the creation (set) or destruction (reset) of a conductive filament that is formed by the random movement of ions in a dielectric. The result of this randomness is a sample of current-voltage curves corresponding to reset-set cycles with variability. So, the variability turns into different voltages and currents within the set and reset processes for each cycle. The analysis of the statistics of the RS operation is essential to understand the devices underlying physics \cite{Acal19,Roldan19,Perez19}. It is necessary to theoretically investigate the stochastic characteristics of RRAMs (directly related to variability), from both the mathematical point of view and the compact modeling perspective. The variability characterization will be essential to develop the infrastructure for device and circuit design software tools.

As the experimental data set connected to each device is a group of current/voltage curves associated with the reset/set cycles of the device, functional data analysis (FDA) methodologies could be the ideal tool to explain the associated variability. Nowadays, FDA is a leading research topic in statistics in which the methods developed for samples of vectors are becoming extended to the case of samples of curves. Besides, the interest in methodological developments as well as in applications to fields such as life science, chemometrics, environment, economy, electronics, among others, are growing continuously. A good review of the main FDA methods, interesting applications and computational algorithms with the free software R can be seen in the books \cite{Ramsay02,Ramsay05,Ramsay09}. The sample curves are usually observed at a finite set of discrete points so that the first step in FDA is usually the smoothing of each sample curve through its representation as a linear combination of basis functions. A comparison of different types of penalized smoothing with B-splines basis was performed in \cite{Aguilera13a}.

The basic tool in FDA is functional principal component analysis (FPCA) that reduces the dimension of the stochastic process generating the sample curves by providing a small set of uncorrelated scalar variables that represent the most important variation modes in the sample. Different penalized PCA approaches for B-spline expansions of smooth functional data were introduced in \cite{Aguilera13b}. FPCA was recently applied to model the variability of the reset processes associated with RRAM devices \cite{Aguilera-Morillo19}.
In the FDA context, the problem  in  the present work consists of  testing homogeneity for several independent samples of experimental data obtained from different RRAMs. The aim is to characterize the device variability by considering different metals as electrode materials and dielectrics of different thicknesses in the fabrication process. The homogeneity problem addressed in this contribution consists of deciding if several independent samples of curves have been generated by the same stochastic process (homogeneity), so that they have equal probability distributions. In order to solve this problem, different parametric and non-parametric approaches based on basis expansion of the sample curves are proposed here.

In the case of normality distribution of the stochastic processes associated with each sample, the homogeneity problem is known as the multi-sample problem or one-way ANOVA problem for functional data, and it is equivalent to equality of the mean functions among the different samples (FANOVA). A detailed description and comparison of tests for the one-way ANOVA problem for functional data can be seen in \cite{Gorecki15,Zhang2014}. Taking into account the basis expansion of the sample curves, the FANOVA is reduced to a multivariate ANOVA (MANOVA) with the vector of basis coefficients of the sample curves as dependent variable and the categorical variable representing the groups as independent variable. The problem is that the current/voltage curves associated with RRAMs are not generated by a Gaussian process so that a different approach is necessary for testing homogeneity. Multivariate non-parametric homogeneity tests \cite{Oja10} on the vector of basis coefficients are considered in this paper to solve the problem. Other important problem is that multivariate homogeneity tests do not perform well with high-dimensional vectors and the number of basis functions needed for an accurate approximation of the sample curves is usually high. In order to solve it, a new approach based on dimension reduction by using FPCA of the sample curves and testing homogeneity on the vector of the most explicative principal components scores is introduced.

Apart from this introduction, the manuscript scheme consists of a theoretical development of functional homogeneity test procedures adapted to the data measured for the devices under study (Section 2), a simulation study to evaluate the performance of the testing approaches in Section 3, an application  with data from resistive memories and the corresponding discussion in Section 4, and finally, the main conclusions in Section 5.

\section{Statistical homogeneity tests for basis expansion of functional data}

Let $\{ x_{ij}(t):  i=1, \dots, m; j=1, \dots, n_i; t\in T \}$  denote $m$ independent samples (groups)
 of curves defined on a continuous
 interval T.  Let us assume that they are realizations of i.i.d. stochastic  processes (functional variables)  $\{ X_{ij}(t): i=1, \dots, m; j=1, \dots, n_i t\in T \}$ with distribution
 $SP(\mu_i (t), \gamma_i (s,t) ), \forall i = 1, . . . , m,$ with $\mu_i (t)$ being the mean function and $\gamma_i (s,t)$ the covariance function associated with each of the $m$ stochastic processes.   Let  us also assume that all sample curves  belong to the Hilbert space L$^2[T]$ of the square
integrable  functions
on $T$, with the natural inner product defined by
$$
<f|g> = \int_T f(t)g(t)dt \ \  \mbox{for all} \ \ f,g\in L^2[T].
$$
The homogeneity of the $m$ samples of curves means that they have been generated by the same stochastic process $SP(\mu (t), \gamma (s,t) )$ with the same probability distribution $\forall i = 1, . . . , m.$ This problem has been recently considered from different points of view. If the processes are Gaussian, then the problem is known as the multi-sample problem or one-way ANOVA problem for functional data (see the book \cite{Zhang2014} for a detailed study). A  comprehensive comparison of tests  for the one-way ANOVA problem for functional data was developed in \cite{Gorecki15}.  More recently, an approach based on the concept of functional depth measures was introduced in \cite{Flores2018}. In this paper we focus on basis expansion of functional data and propose two different type of approaches. One consists on testing multivariate homogeneity on the random vector  of basis coefficients for the $m$ groups, and the other is based on testing multivariate  homogeneity on the associated functional principal components (p.c.'s).

Then, the starting point is to assume that the sample curves  belong to a
finite-dimension space spanned by a basis $\left\{ \phi_{1}\left(
t\right) ,\ldots,\phi_{p}\left(t\right)  \right\},$ so that each stochastic process is represented by its vector  of basis  coefficients. Let us assume that
\begin{equation}
X_{ij} (t) = \sum_{k=1}^p a_{ijk} \phi_k (t) \quad i=1, \dots,m; j=1,\dots, n_i,
\label{bex}
\end{equation}
where $a_{ijk}$ are scalar random variables  with finite variance and $p$ is sufficiently
large to assure an accurate representation of each process. In vector form $X_{ij} (t) = {\bf a}'_{ij}\Phi(t)$ with ${\bf a}_{ij}=(a_{ij1},...,a_{ijp})'$ being the vectors of basis coefficients and $\Phi(t)=(\phi_1(t),...,\phi_p(t))'.$ On the one hand, the selection of the type and dimension of the basis (Fourier, B-splines, wavelets, polinomials, etc) is an important problem that must be solved by taking into account the sample curve characteristics. In the application in this paper a base of cubic splines is chosen because the analysed current/voltage curves are smooth enough. Other useful basis systems are Fourier functions for periodic data, piecewise constant functions for counting processes or wavelets bases for curves with strong local behavior. On the other hand, the basis coefficients are usually estimated by least squares (with or without penalization) from discrete-time noisy observations. A good review about different ways to proceed and how to do it with the software R can be studied in the books \cite{Ramsay05,Ramsay09}.

\subsection{Homogeneity testing on basis coefficients}

The first type considered approach consists of  performing  a  multivariate homogeneity test  on the $m$ samples of the basis coefficient vector $\{{\bf a}_{ij}: i=1, \dots, m; j=1, \dots, n_i\}.$

 When the processes are Gaussian, the one-way ANOVA problem for functional  data is equivalent to equality of the mean functions among the different samples provided that the covariance functions in the groups are equal (homoscedastic case) or different (heteroscedastic case). This problem can be formulated as the hypothesis test of equality of the unknown group mean functions of the $m$ samples
\begin{equation}
H_0 : \mu_1(t) = \dots = \mu_m (t), \forall t\in T,
\label{Fanova1}
\end{equation}
against the alternative that its negation holds.

In the case of the one-way FANOVA problem (\ref{Fanova1}) the functional data verify the following linear model:
\begin{equation}
X_{ij} (t)=\mu(t)+\alpha_i (t)+\epsilon_{ij}(t), \ i=1,...,m, \ j=1,...,n_i,
\label{Fanova3}
\end{equation}
where $\mu (t)$ is the overall mean function, $\alpha_i (t)$ is the {\it i-th} main-effect function, and
 $\epsilon_{ij}(t)$ are the subject-effect functions (i.i.d. errors) with distribution $SP(0, \gamma(s,t) )$ $\forall i=1, \dots,m; j=1,\dots, n_i,$ and $\gamma(s,t)$ being  the common covariance function in the homoscedastic case.

The main-effect functions  are not identifiable so that  in order to be estimated  some constraint must be imposed. The  most used constraint is $\sum_{i=1}^m  \alpha_i(t)=0.$ Under this constraint you have that $\mu_i(t) = \mu(t)+\alpha_i (t).$ Then, by assuming the basis expansion in \ref{bex}, the unbiased estimators of the functional parameters in model \ref{Fanova3} are given by
\begin{itemize}
	\item $\hat{\mu}(t)=\overline{x}(t) = \overline{\bf a}'\Phi(t),$
	\item $\hat{\alpha}_i (t) = \bar{x}_i (t) -\bar{x} (t) = \left ( \overline{\bf a}_i'-\overline{\bf a}' \right ) \Phi(t),$
	\item $\hat{\epsilon}_{ij}(t)=x_{ij}(t)-\overline{x}_i(t) = \left ( {\bf a}_{ij}'-\overline{\bf a}_i' \right ) \Phi(t),$
\end{itemize}
where $\bar{x}$ and $\bar{x}_i (t)$ are the usual  unbiased estimators of the grand mean function and the group mean functions, respectively, and, $\bar{\bf a}$ and $\overline{\bf a}_i$ are the corresponding unbiased estimators of the grand mean vector and the group mean vector associated with the coefficient vectors ${\bf a}_{ij}.$

Taking into account the basis expansions of the sample curves the FANOVA testing problem  is equivalent to the usual multivariate ANOVA test (MANOVA)
 for the matrix of basis coefficients  $A= \left (  a_{(ij)k} \right )_{n\times p},$ with $n =\sum_{i=1}^m n_i.$ This is equivalent to test the equality of mean vectors for the basis coefficients in the $m$ groups. This problem is solved by using one of the well known MANOVA tests: the Wilks's lambda, the Lawley-Hotelling's trace, the Pillai's
trace, and the Roy's maximum root.  In most cases, the exact null distributions of these four test criteria
can not be computed, and approximate F-tests statistics are often used in computer programs. A detailed explanation on these tests can be seen in \cite{Rencher12}.

The main requirements for estimating a one-way  MANOVA model are: 1) observations are randomly and independently sampled from the population; 2) the sample size in each group  must be larger than the number of dependent variables; 3) dependent variables are multivariate normally distributed within each group; 4) homogeneity of variance-covariance matrices in the $m$  groups; and 5) no multicollinearity.

When the $m$  samples of vectors coefficients  are not Gaussian, the F-type
tests described earlier can  not be applicable. Other approaches based on   bootstrap versions of these tests methods may be considered \cite{Gorecki15}. In this paper non-parametric multivariate homogeneity tests based  are considered to solve this problem. Specifically,  the extensions of the univariate Kruskal Wallis's test and  Moods's test \cite{Oja10,Ellis17} that try to check whether the medians are equal in all groups, are applied in the simulation and the application developed in Section 3 and 4, respectively. The Moods's test is less sensitive with the outliers than the Kruskall Wallis's test but it is less powerful when the data are generated from some distributions as, for instance,  the normal distribution.

Let us finally observe that unbalanced sample sizes can lead to unequal variances between samples that could affect to the statistical power and type I error rates of parametric (ANOVA type) tests  \cite{Rusticus14}. In fact, equal sample sizes maximize statistical power. On the other hand,   nonparametric rank-based tests  could lead to paradox results due the non-centralities of the test statistics which may be non-zero for the traditional tests in unbalanced designs. A simple solution is the use of pseudo-ranks instead of ranks \cite{Brunner18}.

\subsection{Homogeneity testing on functional principal components}

This new approach for solving the homogeneity problem with functional data consists of
 reducing the infinite dimension of the stochastic procecess by using FPCA and then performing  a  multivariate homogeneity test on the vectors of the most explicative principal components scores.

FPCA provides the following orthogonal decomposition of the process (Karhunen-Lo\`eve expansion):
\begin{equation}
X_{ij} (t)  = \mu (t) + \sum_{k=1}^\infty f_k (t) \xi_{ijk} ,
\label{K-L}
\end{equation}
where  $\{f_k \}$  are  the  orthonormal  eigenfunctions  of
the covariance operator  associated with its decreasing sequence of non null eigenvalues
$\{\lambda_k \}$, and $\{\xi_{k} \}$ are  uncorrelated zero-mean  random  variables
 (principal components) defined by
$$
\xi_{ijk} =\int_T f_k (t)(X_{ij}(t) -\mu(t))dt.
$$
 The {\it k-th} p.c. $\xi_k$  has  the maximum  variance  $\lambda_k$ out of all  the  generalized   linear
combinations of the functional variable  which are uncorrelated with $\xi_l$  $(l=1,..,k-1).$

By truncating the expression (\ref{K-L}), the process admits a principal component reconstruction
in terms of the first {\it q} principal components so that the sum of their explained variances
is  as  close  as  possible to one. Then, in vector form the functional variable $X(t)$ is
approximated by
$X^q_{ij} (t) - \mu (t) = {\bf \xi}'_{ij} {\bf f }(t),$ with ${\bf \xi}_{ij}=(\xi_{ij1},...,\xi_{ijq})'$ being the vectors of principal components scores  and ${\bf f(t)}=({\bf f}_1(t),...,{\bf f}_q(t))'.$

In practice, and assuming the basis expansion of sample curves given in \ref{bex}, the functional PCA is  equivalent to  multivariate PCA of matrix
$A\Psi^{\frac{1}{2}}$ \cite{Ocana2007}, with $\Psi^{\frac{1}{2}}$ being the squared
root of the matrix of inner products between basis functions $\Psi =\left(\Psi_{ij}\right)_{p \times p} =\int_T \phi_i
\left(t \right) \phi_j \left(t \right) du.$  Then, the
principal component weight function $\hat{f}_k$ admits the basis expansion
$\hat{f}_k \left(t \right)={\bf b}_k' \Phi(t),$
so that,  the vector ${\bf b}_k$ of basis coefficients   is given by ${\bf b}_k = \Psi^{-\frac{1}{2}} {\bf u}_k,$ where
the vectors ${\bf u}_k$ are computed as the solutions to the eigenvalue
problem
$
n^{-1}\Psi^{\frac{1}{2}}A'A\Psi^{\frac{1}{2}}u_k=\lambda_k u_k,
$
where $n^{-1}\Psi^{\frac{1}{2}}A'A\Psi^{\frac{1}{2}}$ is the sample
covariance matrix of $A\Psi^{\frac{1}{2}}.$

Again, we propose two different ways to solve the problem of homogeneity of the vector of the first $q$  principal components in the $m$ groups.
In the case of multivariate normality of the vector of principal components scores, a MANOVA testing procedure based on the F-type statistics is not advisable because the dependent variables are uncorrelated. In this case, we propose to perform  univariate ANOVA on each p.c. score that has more power than MANOVA analysis. In order  to control the Type I error
when conducting these multiple  ANOVA tests, the additive Bonferroni inequality  will be applied so that the alpha level for each ANOVA test is given by the overall level divided by the number of tests. On the other hand, if normality is not verified, then non-parametric   multivariate tests will be applied.

\section{Simulation study}

In this section, an extensive simulation study with artificial data is developed to check the performance of the two functional homogeneity  approaches: one is based on testing homogeneity on the basis coefficients and the other on testing homogeneity on the principal components.

In this study, three groups  have been considered (m=3) with the following three different models for the mean functions:
\begin{itemize}
\item $M 1:$ $\mu_i (t) = 0.1 |\sin {(4\pi t)}| \quad i=1,2,3,$
\item $M 2:$ $\mu_i (t) = 0.05 i |\sin {(4\pi t)}| \quad i=1,2,3,$
\item $M 3:$ $\mu_i (t) = 0.025 i |\sin {(4\pi t)}| \quad i=1,2,3.$
\end{itemize}
Let us observe that $M 1$ corresponds to situations where $H_0$ is true while $M 2$ and $M3$ corresponds to situations where $H_0$ is false. In $M 3$ the differences between the means are smaller so that the testing problem is more difficult.

In addition, two different type of error functions are added to simulate a sample of functional data in the interval [0,1] for each case according to the model in Equation \ref{Fanova3}. For the parametric approaches (Gaussian case), an approximation of the standard Wiener process  given by its Karhunen-Loève expansion truncated in the $qth$ term is used. This is a Gaussian process with covariance function given by $C(t,s) = \sigma^2 \min{(t,s)}.$ The Karhunen-Loeve expansion of this  process is given as follows in terms of the eigenvalues and eigenfunctions of its covariance operator:
$
\epsilon (t)= \sum_{k=1}^\infty \sqrt{\lambda_k} \xi_{k} f_k (t),
$
where the p.c.'s $\xi_{k}$ are independent Gaussian random variables with mean zero and variance one, the eigenvalues are given by
 $\lambda_k = \frac{\sigma^2}{(k-\frac{1}{2})^2 \pi^2},$ with the associated eigenfunctions $f_k (t) = \sqrt{2} \sin{\left ( \left  (k-\frac{1}{2}  \right ) \pi t \right )}.$
The truncation point  in this study is $q=20,$  and five different values for the dispersion parameter $(\sigma = 0.02, \sigma = 0.05, \sigma = 0.10, \sigma = 0.20, \sigma = 0.40)$ are considered. For the non-parametric approaches, the error functions are computed in the same form as the exponential,
adequately centered, of $\epsilon (t)$ (log-normal distribution).

Then, i.i.d. samples, with three different sample sizes $(n_i= 15, n_i = 25, n_i= 35; i=1,2,3),$   are simulated at $51$ equally spaced time points in the interval $[0,1]$ for each one of the thirty considered functional models. Finally, 1000 Monte Carlo replications are developed for each one of the ninety considered cases (three mean models*two type of error *five dispersion parameters*three sample sizes). In order to obtain the basis coefficients for each sample curve from its discretized values in the interval [0,1], least squares approximation in terms of a basis of cubic B-splines of dimension 18 was used in all cases. All the computations  were obtained with the packages {\it fda} \cite{Ramsay09} and {\it npmv} \cite{Ellis17} of statistical software  {\it R.} As indicator of the test
performance, the observed acceptance proportions at a significance level 0.05 under every considered model were computed. Three different number of p.c.'s were considered for the principal component approach: the first three p.c.'s,  the first five p.c.'s and the first eight p.c.'s that explain approximately a $95 \%,$ a $97 \%,$  and a $99 \%$ of the total variability, respectively. The results  for the two testing parametric approaches  with the F-type tests (Gaussian errors) appear in Table \ref{Table1}. MANOVA testing with the Pillai statistics was conducted for the basis coefficients approach and multiple univariate ANOVA for the principal component approach, using Bonferroni's inequality for preserving the overall significance level. On the other hand, the results for the non-parametric approaches (Log-normal errors) appear in Table \ref{Table2}. The multivariate extension of the  Kruskall-Wallis univariate test was used to compute the p-values.

Next, a discussion of the simulation experiment is presented that can help to show the practical utility of the proposed methodology:
 \begin{enumerate}
 \item An important key point to keep in mind is the dispersion parameter $\sigma$. It seems that  the testing performance depends strongly on the error dispersion, getting worse as $\sigma$ increases in all cases. In fact, when $\sigma = 0.40$ the power of the tests is too small, especially in the case of the model M3 in which the differences between the group means are smaller. This  must be taken into account for future analysis because previous simulation studies of this type (see  \cite{Gorecki15}) do not consider a value of $\sigma$ higher than 0.20.
 \item Another interesting point has to do with sample sizes. For small values of $\sigma$ the sample size does not have an important effect in the power of the test. However, the sample size  plays a fundamental role when $\sigma$ increases, as the test converts into less conservative for the cases M2 and M3.
\item Regarding to the number of p.c.'s selected for the testing procedure, it can be  seen in Tables \ref{Table1} and \ref{Table2} that the greater the number of p.c.'s, the better results the tests achieve. In fact, the tests don't behave well in the situations where the variability explained is lower than $99 \%$ and the term $\sigma$ is large. So, it would be recommendable to consider a number of p.c.'s that guarantees  around the $99 \%$ of the variability.
\item For the model M1 ($H_0$ is true), both the parametric and the non-parametric tests provide excellent results. The acceptance proportions are greater than 0.938 in all the cases.
\item In the case of model M2, the results obtained in the parametric case with the basis coefficients  and  with  eight p.c.'s  are really good, even when the dispersion is very high $(\sigma=0.4).$  Only some problems are detected when the sample size is small in this case. Nevertheless, the tests provide slightly better results for the basis coefficients approach. On the other hand, the outputs for M2 when we consider the non-parametric tests change a bit in comparison with the previous situation. Now, the basis coefficients model does not work very well when the sample is not large enough for $\sigma=0.20$ and for $\sigma=0.40$, with the acceptance proportion being 0.169 and 0.706, respectively. Instead, if we consider the approach with 8 p.c.'s the results are much better, only having controversy when $n_i=15$ and $\sigma=0.40$ just like in the parametric case.
\item The behavior of results with  model M3 are very similar to the case of model  M2, basis coefficients approach is slightly better in the parametric case but it occurs the opposite in the non-parametric case. In addition, the tables bring to light the lack of power of both tests (parametric and non-parametric) when the differences among group means are small and $\sigma$ is large. We are rather concerned with the frequency of a correct decision in these situations.
\end{enumerate}

To sum  up, we can firstly conclude that the parametric tests are more powerful than the non-parametric ones and, for that reason, they must be considered when the conditions of validity are satisfied. Another important aspect to keep in mind is the reduction of the dimension provided by the principal component approach. This approach can be really interesting when the number of dependent variables (basis coefficients) is large and  the problem is reduced to testing homogeneity on a small number of p.c's. Based on the results of this simulation study, it can be concluded that the principal component approach explaining a $99\%$ of variability  gives  better results than the basis coefficients approach in the non-parametric case. Regarding to the parametric case, the fact of using the Bonferroni's inequality for correcting the  significance level in the multiple ANOVA tests on the p.c.'s, could be the reason for a slight  decrease in the power with respect to the basis coefficient approach in this study where $8$ p.c.'s are needed to explain at least a $99 \%$ of the total variability. This problem disappears in practice when  only two or three components are necessary so that the corrected level of significance is not so small and the acceptance proportion would increase.

\begin{table}
	\centering
	\resizebox{10cm}{!} {
	\begin{tabular}{l|l|llllll}
		\hline
		Mean                & $n_i$                  & Model       & $\sigma$=0.02 & $\sigma$=0.05 & $\sigma$=0.10 & $\sigma$=0.20 & $\sigma$=0.40  \\ \hline
		\multirow{12}{*}{M1} & \multirow{4}{*}{15} & Basis coef. & 0.949  & 0.956  & 0.951  & 0.957  & 0.944   \\
		&                     & 3 p.c.'s  & 0.938  & 0.954  & 0.968  & 0.959  & 0.949   \\
		&                     & 5 p.c.'s   & 0.948  & 0.944  & 0.941  & 0.945  & 0.958   \\
		&                     & 8 p.c.'s   & 0.958  & 0.955  & 0.946  & 0.946  & 0.942   \\ \cline{2-8}
		& \multirow{4}{*}{25} & Basis coef. & 0.952  & 0.953  & 0.953  & 0.964  & 0.951   \\
		&                     & 3 p.c.'s   & 0.952  & 0.942  & 0.953  & 0.948  & 0.959   \\
		&                     & 5 p.c.'s   & 0.954  & 0.962  & 0.945  & 0.948  & 0.950   \\
		&                     & 8 p.c.'s   & 0.962  & 0.947  & 0.951  & 0.954  & 0.956   \\ \cline{2-8}
		& \multirow{4}{*}{35} & Basis coef. & 0.949  & 0.953  & 0.944  & 0.948  & 0.946   \\
		&                     & 3 p.c.'s   & 0.951  & 0.956  & 0.951  & 0.952  & 0.959   \\
		&                     & 5 p.c.'s   & 0.949  & 0.953  & 0.954  & 0.953  & 0.948   \\
		&                     & 8 p.c.'s   & 0.951  & 0.955  & 0.960  & 0.943  & 0.957   \\ \hline
		\multirow{12}{*}{M2} & \multirow{4}{*}{15} & Basis coef. & 0      & 0      & 0      & 0      & 0.170    \\
		&                     & 3 p.c.'s   & 0      & 0      & 0.001  & 0.448  & 0.852   \\
		&                     & 5 p.c.'s   & 0      & 0      & 0      & 0.043  & 0.681   \\
		&                     & 8 p.c.'s   & 0      & 0      & 0      & 0.003  & 0.298   \\ \cline{2-8}
		& \multirow{4}{*}{25} & Basis coef. & 0      & 0      & 0      & 0      & 0.005   \\
		&                     & 3 p.c.'s   & 0      & 0      & 0      & 0.265  & 0.789   \\
		&                     & 5 p.c.'s   & 0      & 0      & 0      & 0.006  & 0.595   \\
		&                     & 8 p.c.'s   & 0      & 0      & 0      & 0      & 0.042   \\ \cline{2-8}
		& \multirow{4}{*}{35} & Basis coef. & 0      & 0      & 0      & 0      & 0       \\
		&                     & 3 p.c.'s   & 0      & 0      & 0      & 0.157  & 0.746   \\
		&                     & 5 p.c.'s   & 0      & 0      & 0      & 0.001  & 0.552   \\
		&                     & 8 p.c.'s   & 0      & 0      & 0      & 0      & 0.007   \\ \hline
		\multirow{12}{*}{M3} & \multirow{4}{*}{15} & Basis coef. & 0      & 0      & 0      & 0.181  & 0.783   \\
		&                     & 3 p.c.'s   & 0      & 0      & 0.465  & 0.879  & 0.937   \\
		&                     & 5 p.c.'s   & 0      & 0      & 0.043  & 0.724  & 0.914   \\
		&                     & 8 p.c.'s   & 0      & 0      & 0.004  & 0.307  & 0.752   \\ \cline{2-8}
		& \multirow{4}{*}{25} & Basis coef. & 0      & 0      & 0      & 0.005  & 0.537   \\
		&                     & 3 p.c.'s   & 0      & 0      & 0.286  & 0.802  & 0.925   \\
		&                     & 5 p.c.'s   & 0      & 0      & 0.009  & 0.607  & 0.879   \\
		&                     & 8 p.c.'s   & 0      & 0      & 0      & 0.044  & 0.627   \\ \cline{2-8}
		& \multirow{4}{*}{35} & Basis coef. & 0      & 0      & 0      & 0      & 0.330   \\
		&                     & 3 p.c.'s   & 0      & 0      & 0.178  & 0.724  & 0.903   \\
		&                     & 5 p.c.'s   & 0      & 0      & 0      & 0.485  & 0.880   \\
		&                     & 8 p.c.'s   & 0      & 0      & 0      & 0.004  & 0.466  \\ \hline
	\end{tabular}
	}
\caption{Observed acceptance proportions for each scenario  at a significance level 0.05 in the case of Gaussian errors.}
\label{Table1}
\end{table}

\newpage

\begin{table}
	\centering
	\resizebox{10cm}{!} {
	\begin{tabular}{l|l|llllll}
		\hline
		Mean                & $n_i$                  & Model       & $\sigma$=0.02 & $\sigma$=0.05 & $\sigma$=0.10 & $\sigma$=0.20 & $\sigma$=0.40  \\ \hline
		\multirow{12}{*}{M1} & \multirow{4}{*}{15} & Basis coef. & 0.984  & 0.983  & 0.975  & 0.982  & 0.982   \\
		&                     & 3 p.c.'s   & 0.960  & 0.951  & 0.957  & 0.946  & 0.960   \\
		&                     & 5 p.c.'s   & 0.967  & 0.965  & 0.950  & 0.950  & 0.951   \\
		&                     & 8 p.c.'s   & 0.966  & 0.972  & 0.952  & 0.959  & 0.970   \\ \cline{2-8}
		& \multirow{4}{*}{25} & Basis coef. & 0.966  & 0.962  & 0.978  & 0.964  & 0.972   \\
		&                     & 3 p.c.'s   & 0.951  & 0.943  & 0.945  & 0.964  & 0.930   \\
		&                     & 5 p.c.'s   & 0.951  & 0.958  & 0.955  & 0.961  & 0.953   \\
		&                     & 8 p.c.'s   & 0.943  & 0.950  & 0.955  & 0.963  & 0.959   \\ \cline{2-8}
		& \multirow{4}{*}{35} & Basis coef. & 0.960  & 0.959  & 0.971  & 0.967  & 0.967   \\
		&                     & 3 p.c.'s   & 0.947  & 0.950  & 0.941  & 0.949  & 0.956   \\
		&                     & 5 p.c.'s   & 0.954  & 0.949  & 0.958  & 0.957  & 0.950   \\
		&                     & 8 p.c.'s   & 0.960  & 0.964  & 0.952  & 0.957  & 0.949   \\ \hline
		\multirow{12}{*}{M2} & \multirow{4}{*}{15} & Basis coef. & 0.005  & 0.009  & 0.011  & 0.169  & 0.706   \\
		&                     & 3 p.c.'s   & 0      & 0      & 0.001  & 0.525  & 0.894   \\
		&                     & 5 p.c.'s   & 0      & 0      & 0      & 0.041  & 0.793   \\
		&                     & 8 p.c.'s   & 0      & 0      & 0      & 0      & 0.365   \\ \cline{2-8}
		& \multirow{4}{*}{25} & Basis coef. & 0      & 0      & 0      & 0      & 0.136   \\
		&                     & 3 p.c.'s   & 0      & 0      & 0      & 0.309  & 0.865   \\
		&                     & 5 p.c.'s   & 0      & 0      & 0      & 0      & 0.663   \\
		&                     & 8 p.c.'s   & 0      & 0      & 0      & 0      & 0.046   \\ \cline{2-8}
		& \multirow{4}{*}{35} & Basis coef. & 0      & 0      & 0      & 0      & 0.010   \\
		&                     & 3 p.c.'s   & 0      & 0      & 0      & 0.18   & 0.795   \\
		&                     & 5 p.c.'s   & 0      & 0      & 0      & 0      & 0.568   \\
		&                     & 8 p.c.'s   & 0      & 0      & 0      & 0      & 0.006   \\ \hline
		\multirow{12}{*}{M3} & \multirow{4}{*}{15} & Basis coef. & 0.007  & 0.025  & 0.142  & 0.697  & 0.940   \\
		&                     & 3 p.c.'s   & 0      & 0.001  & 0.484  & 0.890  & 0.933   \\
		&                     & 5 p.c.'s   & 0      & 0      & 0.044  & 0.775  & 0.910   \\
		&                     & 8 p.c.'s   & 0      & 0      & 0.001  & 0.322  & 0.863   \\ \cline{2-8}
		& \multirow{4}{*}{25} & Basis coef. & 0      & 0      & 0      & 0.112  & 0.784   \\
		&                     & 3 p.c.'s   & 0      & 0      & 0.325  & 0.840  & 0.916   \\
		&                     & 5 p.c.'s   & 0      & 0      & 0      & 0.636  & 0.908   \\
		&                     & 8 p.c.'s   & 0      & 0      & 0      & 0.039  & 0.716   \\ \cline{2-8}
		& \multirow{4}{*}{35} & Basis coef. & 0      & 0      & 0      & 0.006  & 0.606   \\
		&                     & 3 p.c.'s   & 0      & 0      & 0.157  & 0.776  & 0.917   \\
		&                     & 5 p.c.'s   & 0      & 0      & 0      & 0.542  & 0.887   \\
		&                     & 8 p.c.'s   & 0      & 0      & 0      & 0.002  & 0.557  \\ \hline
	\end{tabular}
	}
\caption{Observed acceptance proportions for each scenario  at a significance level 0.05 in the case of Log-normal errors}
\label{Table2}
\end{table}

\section{Application results and discussion}

In this paper, we will use experimental data measured at the Institute of Microelectronics of Barcelona (CNM-CSIC) where the devices were also fabricated. The devices are based on a metal-oxide-semiconductor stack \cite{Gonzalez14}. The metal electrodes employed were Ni and Cu, the dielectrics (Hf$O_2$) and Si-$n^+$ was employed as bottom electrode. In particular, the following devices were used: Device 1 (DV1): Ni/Hf$O_2$ (20 nm thick)/Si-$n^+$, Device 2 (DV2) Ni/{Hf$O_2$(10 nm thick)/Si-$n^+$, Device 3 (DV3): Cu/Hf$O_2$(20 nm thick)/Si-$n^+.$ The I–V characteristics were measured using a HP-4155B semiconductor parameter analyzer. A negative voltage was employed although we used the absolute value for easiness in the numerical analysis. The functional homogeneity approaches presented here will be applied to decide about the existence of significant statistical differences between the three  devices considered.\\

More precisely, RRAM operation is based on the stochastic nature of resistive switching processes; these, in the most cases, create and rupture conductive filaments that change drastically the resistance of the device. These processes are known as set and reset, respectively. Moreover, the resistance change gives rise to a sample of current-voltage curves corresponding to the reset-set cycles, where the mentioned variability is translated to different voltages and currents related to set and reset processes for each cycle. See the set and reset curves in Figure 1 of reference \cite{Aguilera-Morillo19} and the variation of the set and reset voltages, where the current drastically increases or drops off.

In this study, we have information about  2782 reset curves corresponding to the device with the nickel electrode and the dielectric 20 nanometres thick (Device 1), 1742 reset curves for the device with nickel electrode and a dielectric 10 nanometres thick (Device 2)  and 233 reset curves for devices with a copper electrode and a dielectric 20 nanometres thick (Device 3), denoted as $\{ I_{ij}(v): v \in [0,V_{ij-reset}] \}$ being $i=1,2,3$ the type of device and $j=1,...,n_i$ the sample size of the group $i$. It would have been interesting to have at our disposal data related to RRAMs fabricated with a copper electrode and a dielectric 10 nanometres thick, but for reasons connected to the fabrication plans, it was not possible.

From mathematical viewpoint, and before applying FDA, the reset curves require some previous transformations because they are not defined on the same domain (reset voltages are different in each curve due to variability), and we only have discrete observations at a finite set of current values until the reset voltage is achieved in each curve. In order to solve these problems, \cite{Aguilera-Morillo19} proposed a simple FDA approach to analyze these kind of curves prior to apply some specific statistical FDA techniques. Firstly, the initial domain $[0,V_{ij-reset}]$ was transformed in the interval $[0,1]$ in a way that every registered sample curve $I_{ij}^*(u) (u\in [0,1])$ has a new set of arguments given by transformation $u=v/V_{ij-reset}.$ Secondly, taking into account that the curves are smooth enough, P-spline smoothing with B-spline bases was used to reconstruct all reset curves. The principal reasons why P-spline are usually considered  a great accurate approximation of sample curves, are  less numerical complexity and computational cost, and that the choice and position of knots is not determinant, so that it is sufficient to choose a relatively large number of equally spaced basis knots \cite{Aguilera13b}. In this paper, for each reset curve of the three devices, it has been considered a cubic B-spline basis of dimension 20 with 17 equally spaced knots in the interval $[0,1]$ and a penalty parameter $\lambda=0.5$. In order to select the same smoothing parameter for all the sample paths a leave-one-out cross validation procedure was used.

Let us remember that the aim is to test if there are significant differences between RRAMs of the three different technologies under study. The first step is to test the equality of the three unknown mean functions by using the one-way FANOVA approach under the assumption that the reset curves of each group are generated by a Gaussian process with the same covariance operator. The estimation of the sample mean function in each group is displayed in Figure 1 (bottom-right) next to all the corresponding smoothed registered curves. Graphically, it seems that there are differences depending on the type of material and thickness.\\

In order to test the equality of the three unknown mean functions, MANOVA on the matrix of basis coefficients $A_{(4757 \times 20)}$ could be applied (see Subsection 2.1) so that we have $20$ dependent quantitative variables (the dimension  of the B-spline basis) and one independent categorical variable (the three types  of devices). It is well known that the main purpose of this technique is to compare the mean vectors of the three  samples for significant differences. Equality of the mean vectors implies that the three single means are equal for each dependent variable. Before applying MANOVA, we must verify that the vectors of basis coefficients of each device technology have multivariate normal distribution with equal covariance matrices. However, these hypothesis are not fulfilled for the considered reset curves. As a matter of fact, the p-values associated with the Kullback's test or M-Box's test for the homogeneity of covariance matrices, and the p-value linked with the Kolmogorov-Smirnov's test for the univariate normality of each single basis coefficient are all $<0.001.$ This means that the assumptions of multivariate normality and homogeneity of covariance matrices are not true. Therefore, the second step  consists of using a non-parametric test for the homogeneity of the vectors of basis coefficients in the three devices.  In this application,  due to the high presence of outliers, the extension of the univariate Mood's test which is based on spatial signs is employed (see results in Table \ref{Table3}). Taking into account that the associated p-value is less than 0.001, we can conclude that the reset voltage distribution is different according the kind of metal for the electrode and dielectric thickness used in the RRAM technologies. \\

Finally, we are going to test homogeneity on the functional principal components computed from the P-spline smoothing of the sample curves. The percentages of variance explained by the first four principal components are $99.639,$ $0.284,$  $0.046$ and $0.020,$ respectively.
Let us observe that only the first principal component explains more than $99\%$ of the total variability of the process. Hence, by truncating K-L expansion \ref{K-L} the reset process can be represented as $I{^*}^1(u)=\overline{I}^*(u)+\xi_1^*f_1^*(u), \ u \in [0,1],$
where $\xi_1^*$ is an scalar random variable called first principal component score and $ f_1^*$ is a function that represents the principal component weight curve. Thus, the problem of homogeneity is reduced to one-way ANOVA  for the first principal component if this variable is normal distributed and with the same variance in the three devices. However, neither the normality nor the homogeneity of variance are accepted for these data so that the p-values associated with the corresponding tests (Kolmogorov-Smirnov's test for the univariate normality and Levene's Test for homogeneity of variance) are less than 0.001. Again,  the  ANOVA methodology can not be applied and so, non-parametric tests are used in order to test the differences among group means of the first p.c.  Specifically, univariate  Mood’s median test is used for the general homogeneity hypothesis (see results in Table \ref{Table3}).  It could be also interesting to test whether there are differences among pairs of devices in order to prove if the dielectric thickness or the electrode material by separated play some important role in the RRAMs operation. Wilcoxon's rank sum test is applied for the pairwise comparisons by means of the Benjamini’s method for adjusting p-values. In both cases the associated p-values are less than $<0.001.$ On the other hand, the p-values provided by Wilcoxon's rank sum test for the pairwise comparisons are also smaller than $0.001.$ Based on these results we can conclude that the distribution of the first p.c. are  significantly different for the three considered devices. Therefore,
it can be highlighted in what is referred to the reset curves of the technologies under consideration here that the type of metal employed for the electrodes and the dielectric thickness have a high influence on RRAMs operation and in the statistical information linked to their inherent variability.

\section{Conclusions}

The aim of this work is to decide if there are significant differences in the probability distribution  that generates the reset processes associated with RRAMs fabricated making use of different materials for the electrodes and using dielectrics of different thicknesses. From the methodological point of view, this homogeneity  problem consists of  testing if  different samples (groups) of curves come from the same population. Several  FDA approaches  have been proposed in literature when the stochastic processes associated with each sample are Gaussian. This problem is known as multi-sample problem or FANOVA and consists of testing the equality of the  group mean functions. If the normality assumption is not true some bootstrap approaches were developed. In this paper, two different parametric and non-parametric homogeneity testing approaches are proposed by assuming a basis expansion of the sample curves. Both are reduced to testing multivariate homogeneity (parametric and non-parametric), the first one  on a vector basis coefficients and the second one on a vector of principal component scores. The different proposals are motivated by the statistical study of the variability in the three samples of reset curves   analyzed at the end of the paper. On the other hand, an extensive simulation study has been developed to check the practical performance of the testing approaches.  In this study, the influence of sample size and variability of errors has been revealed, in addition to the improvement in the behavior of the tests with the principal component approach for the non-parametric case.

\begin{table}
\begin{center}
	\begin{tabular}{c|ccc}
		\hline
		& Chi-squared & df & p-value \\
		\hline
		Basis coef. & 5516.5 & 40 & $<0.001$ \\
		First p.c. & 2538.2 & 2 & $<0.001$ \\
		\hline
	\end{tabular}
\caption{Chi-squared test statistic, the degrees of freedom of its approximated chi-squared distribution and the p-value  for the  Mood's median test}
\label{Table3}
\end{center}
\end{table}

\section*{Acknowledgements}

We would like to thank F. Campabadal and M. B. González from the IMB-CNM
(CSIC) in Barcelona for fabricating and providing the experimental measurements of
the devices employed here. The authors thank the support of the Spanish Ministry of
Science, Innovation and Universities under projects TEC2017-84321-C4-3-R,
MTM2017-88708-P,  IJCI-2017-34038 (also supported by the FEDER program), project A.TIC.117.UGR18 funded by the government of Andalusia (Spain)
and the FEDER program and the PhD grant (FPU18/01779) awarded to  Christian Acal.
This work has made use of the Spanish ICTS Network MICRONANOFABS.

\begin{figure}
	\centering
	\includegraphics[scale=0.5]{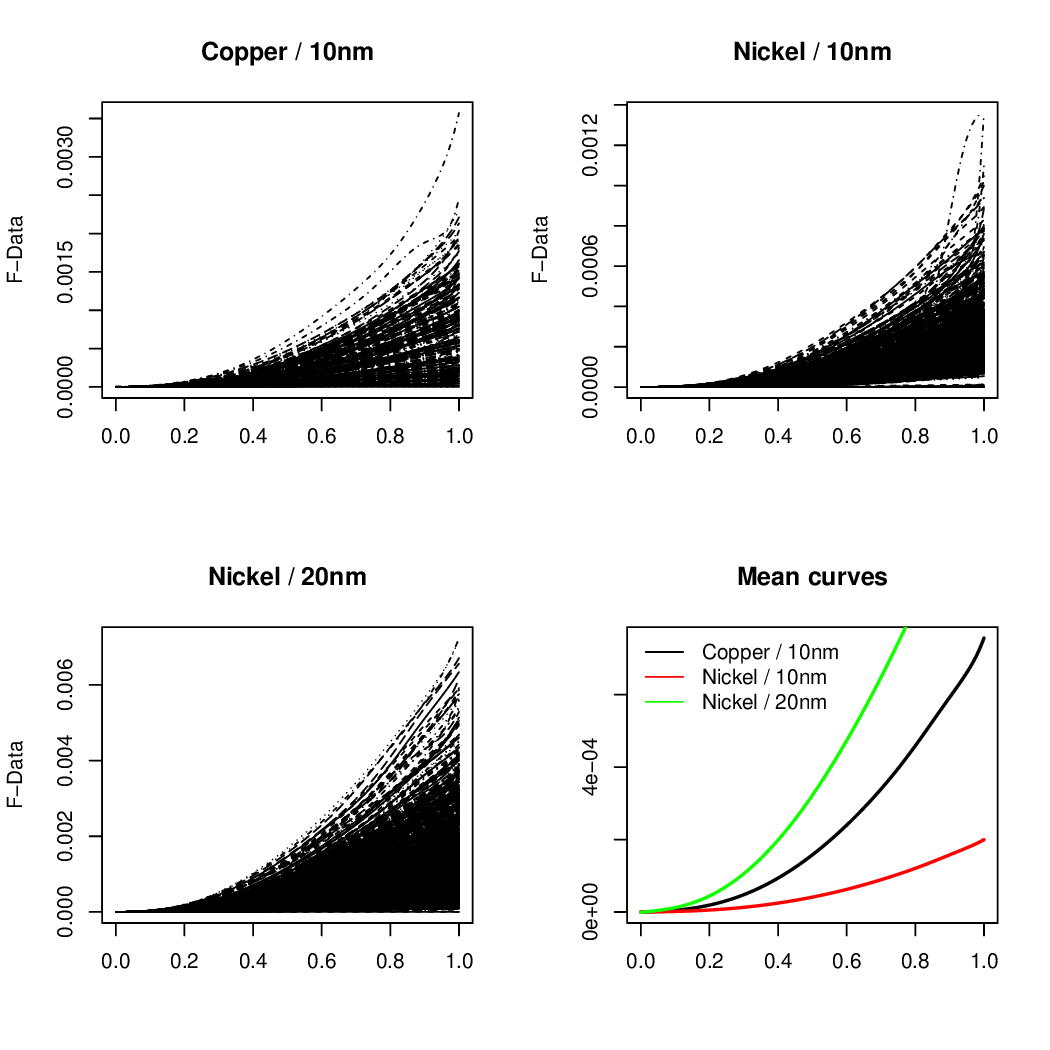}
	\caption{Sample group mean functions (bottom-right) and all the P-spline smoothed registered curves for each type of device. }
\end{figure}


\bibliography{Biblio_engineering_revised}

\end{document}